\documentclass[11pt]{article}
\usepackage{amsfonts}
\usepackage{graphicx}
\usepackage{epstopdf}
\usepackage{epsfig}
\usepackage{amssymb}
\usepackage{setspace}
\usepackage{caption}
\usepackage{amsfonts}
\usepackage{color}
\usepackage{amsmath}
\usepackage{float}
\usepackage{comment}
\usepackage{subcaption}
\newcommand {\be}{\begin{equation}}
\newcommand {\ee}{\end{equation}}
\newcommand {\bea}{\begin{array}}
	
	\newcommand {\eea}{\end{array}}

\textwidth=6.5in \hoffset=-.75in \voffset=-1in \numberwithin{equation}{section}
\numberwithin{figure}{section}

\begin{document}

	\begin{titlepage}
		\vspace{1cm}
		\begin{center}
			{\Large \bf {Kerr-Bertotti-Robinson Spacetime and the Kerr/CFT Correspondence}}\\
		\end{center}
		\vspace{2cm}
		\begin{center}
			\renewcommand{\thefootnote}{\fnsymbol{footnote}}
			Haryanto M. Siahaan{\footnote{haryanto.siahaan@unpar.ac.id}}\\
			Program Studi Fisika, Fakultas Sains,\\	Universitas Katolik Parahyangan,\\
			Jalan Ciumbuleuit 94, Bandung 40141, Indonesia
			\renewcommand{\thefootnote}{\arabic{footnote}}
		\end{center}
		
	\begin{abstract}
We construct the Kerr/CFT correspondence for extremal Kerr--Bertotti--Robinson (Kerr--BR) black holes, which are exact stationary solutions of the Einstein--Maxwell equations describing a rotating black hole immersed in a uniform Bertotti--Robinson electromagnetic universe. After reviewing the geometry, horizon structure, and thermodynamics of the Kerr--BR family, we demonstrate that the external field non-trivially modifies both the horizon positions and the extremality condition. For extremal configurations, the near-horizon limit yields a warped $\mathrm{AdS}_3$ geometry with an associated Maxwell field aligned to the $U(1)$ fibration. Imposing standard Kerr/CFT boundary conditions, the asymptotic symmetry algebra gives rise to a Virasoro algebra with central charge $c_L$ and left-moving temperature $T_L$ that depend explicitly on the external field strength. The Cardy formula then reproduces exactly the Bekenstein--Hawking entropy of the extremal Kerr--BR black hole for any admissible value of the Bertotti--Robinson field, thereby establishing a consistent Kerr/CFT dual description. Comparisons with the magnetized Melvin--Kerr and Kaluza--Klein black holes are briefly discussed, highlighting qualitative differences in their curvature profiles and horizon geometries.
	\end{abstract}
	\end{titlepage}\onecolumn
	\bigskip

\section{Introduction}
\label{sec:intro}

Black holes in regimes of strong gravity are central to both astrophysics and theoretical physics. They are now firmly established as real objects, from stellar-mass systems in X-ray binaries~\cite{Remillard:2006fc} to the supermassive black holes in galactic centres~\cite{Kormendy:2013dxa}. Theoretically, their description is dominated by a small set of exact solutions: Schwarzschild~\cite{Schwarzschild:1916uq}, Reissner--Nordstr\"om~\cite{Reissner:1916cle}, Kerr~\cite{Kerr:1963ud}, and Kerr--Newman~\cite{Newman:1965my}. These spacetimes are stationary, axisymmetric, asymptotically flat, and of Petrov type~D~\cite{Debever:1971zz}, with aligned gravitational and electromagnetic principal null directions---properties that make their geodesic and wave dynamics highly tractable~\cite{Chandrasekhar:1985kt}.

A natural extension, motivated by astrophysical environments, is to consider black holes embedded in external electromagnetic fields. Such models are useful for studying objects interacting with magnetized accretion flows and jets~\cite{Nishikawa:2004wp,Cheong:2024stz}. A standard approach uses the Melvin magnetic universe~\cite{Melvin1964,Melvin:1965zza}. Applying the Harrison transformation~\cite{Harrison:1968zz} yields solutions like the Kerr--Melvin black hole~\cite{Ernst1976,Ernst:1976mzr}, which have been used to study energy extraction~\cite{Bicak:1985rw} and magnetohydrodynamics. However, these spacetimes have global limitations: the magnetic field is cylindrically focused, decays at infinity, and often introduces conical singularities~\cite{Booth:2015nwa}.

An alternative is to immerse a black hole in a Bertotti--Robinson (BR) universe~\cite{Bertotti:1959pf,Robinson:1959ev}. This spacetime, a direct product $\mathrm{AdS}_2\times S^2$ supported by a homogeneous electromagnetic field, represents the near-horizon geometry of an extremal Reissner--Nordstr\"om black hole~\cite{Clement:2001ny}. Recently, Podolsk{\'y} and Ovcharenko constructed the Kerr--Bertotti--Robinson (Kerr--BR) solution~\cite{Podolsky:2025tle,Ovcharenko:2025cpm}: a three-parameter family $(M,a,B)$ describing a Kerr black hole in a uniform BR field. It reduces to Kerr for $B\to0$ and to a (conformal) BR universe for $M\to0$. A key feature, distinct from the Pleba{\'n}ski--Demia{\'n}ski~\cite{Plebanski:1976gy} and Kerr--Melvin classes, is the non-alignment of the Maxwell and Weyl tensor principal null directions, despite the spacetime remaining algebraically special. Subsequent work has explored its hidden symmetries~\cite{Gray:2025lwy}, optical properties~\cite{Zeng:2025tji}, test particle dynamics~\cite{Zhang:2025ole}, embeddings in non-linear electrodynamics~\cite{Ortaggio:2025sip} and Melvin-Bonnor universe \cite{Astorino:2025lih}, establishing Kerr--BR as a fertile ground for strong-field studies.

In parallel, the Kerr/CFT correspondence has provided a powerful microscopic perspective on extremal black holes. Guica \emph{et al.}~\cite{Guica:2008mu} showed that the near-horizon geometry of an extremal Kerr black hole (NHEK) possesses a warped $\mathrm{AdS}_3$ structure whose asymptotic symmetries yield a chiral Virasoro algebra. Identifying the black hole with a thermal state in a dual CFT with Frolov--Thorne temperature $T_L$~\cite{Frolov:1989jh} allows its Bekenstein--Hawking entropy to be reproduced via the Cardy formula~\cite{Cardy:1986ie}. This framework has been successfully extended to numerous charged, accelerating, and magnetized black holes~\cite{Hartman:2008pb,Azeyanagi:2008kb,Compere:2012jk,Astorino:2015naa,Siahaan:2015xia,Siahaan:2021uqo,Ghezelbash:2021xvc,Siahaan:2024zhx}.

This paper bridges these two lines of inquiry by constructing the Kerr/CFT correspondence for extremal Kerr--BR black holes. The global structure of Kerr--BR---asymptotically $\mathrm{AdS}_2\times S^2$ rather than flat or Melvin---poses a novel context for holography. We demonstrate that its extremal near-horizon geometry retains a warped $\mathrm{AdS}_3$ structure with an $SL(2,\mathbb{R})\times U(1)$ isometry. Imposing Kerr/CFT boundary conditions, we derive the central charge $c_L$ and left-moving temperature $T_L$ of the dual chiral CFT, which depend explicitly on the external field strength $B$. The Cardy formula then yields an entropy that matches the Bekenstein--Hawking result exactly for all admissible $B$, confirming the robustness of the Kerr/CFT paradigm in this homogeneous electromagnetic background. As a complementary analysis, we also compute and compare the curvature profiles (Kretschmann scalar) of Kerr--BR and Melvin--Kerr spacetimes, highlighting distinctive geometric features induced by the different external fields.

The paper is organized as follows. Section~\ref{sec:KBR-review} reviews the Kerr--BR geometry, horizon structure, and thermodynamics. Section~\ref{sec:NHEKBR} derives the near-horizon limit of extremal Kerr--BR black holes, computes the asymptotic symmetry algebra and central charge, and verifies the entropy matching via the Cardy formula. Finally we conclude with a summary and outlook. Explicit expressions for the equatorial squared Riemann tensor in both Kerr--BR and Melvin--Kerr spacetimes are provided in an Appendix.

\section{Kerr--Bertotti--Robinson geometry}
\label{sec:KBR-review}

In this section, we briefly summarize the Kerr--Bertotti--Robinson (Kerr--BR) spacetime constructed in Ref.~\cite{Podolsky:2025tle}. This solution describes a rotating black hole immersed in a uniform Bertotti--Robinson electromagnetic universe; therefore, the material presented here closely follows Ref.~\cite{Podolsky:2025tle} and is required for our aim of constructing the Kerr/CFT correspondence for extremal Kerr--BR black holes. We work in Boyer--Lindquist--type coordinates $(t,r,x=\cos\theta,\phi)$ and use the notation $M$ for the black-hole mass, $a$ for the rotation parameter, and $B$ for the external field strength.

The line element of Kerr--BR spacetime can be written in the compact form \cite{Podolsky:2025tle}
\begin{equation}
{\rm d}s^2 = \frac{1}{\Omega^2}
\bigg[
- \frac{Q}{\rho^2} \bigl({\rm d}t - a \Delta_x\, {\rm d}\phi\bigr)^2
+ \frac{\rho^2}{Q}\, {\rm d}r^2
+ \frac{\rho^2}{P\Delta_x}\, {\rm d}x^2
+ \frac{P\Delta_x}{\rho^2} \,
\bigl(a\,{\rm d}t - (r^2 + a^2)\,{\rm d}\phi\bigr)^2
\bigg],
\label{eq:KBR-metric}
\end{equation}
where the metric functions are
\begin{align}
\rho^2 &= r^2 + a^2 x^2,
\label{eq:rhoKBR}
\\[2pt]
P &= 1 + B^2\Bigl(M^2 \frac{I_2}{I_1^2} - a^2\Bigr) x^2,
\label{eq:PKBR}
\\[2pt]
Q &= (1 + B^2 r^2)\,\Delta,
\label{eq:QKBR}
\\[2pt]
\Omega^2 &= (1 + B^2 r^2) - B^2 \Delta x^2,
\label{eq:OmegaKBR}
\\[2pt]
\Delta &= \Bigl(1 - B^2 M^2 \frac{I_2}{I_1^2}\Bigr) r^2 - 2M \frac{I_2}{I_1} r + a^2,
\label{eq:DeltaKBR}
\end{align}
and
\begin{equation}
I_1 = 1 - \tfrac{1}{2} B^2 a^2,
\qquad
I_2 = 1 - B^2 a^2.
\label{eq:I1I2KBR}
\end{equation}

The electromagnetic field is conveniently encoded in a complex 1-form potential
\begin{equation}
A_\mu {\rm d}x^\mu = \frac{e^{i\sigma}}{2B}
\bigg[
\Omega_r\,
\frac{a\,{\rm d}t - (r^2 + a^2)\,{\rm d}\phi}{r + i a x}
- i \Omega_x\,
\frac{{\rm d}t - a \Delta_x\, {\rm d}\phi}{r + i a x}
+ (\Omega - 1)\, {\rm d}\phi
\bigg],
\label{eq:KBR-potential}
\end{equation}
where \(\sigma\) is a duality rotation parameter, and 
\be \label{eq.Omega_r}
\Omega _r  \equiv \frac{{\partial \Omega }}{{\partial r}} = \frac{B^2}{\Omega}\left[
r(1 - x^2)
+ x^2 I_2\left(
\frac{M}{I_1}
+ \frac{B^2 M^2 r}{I_1^2}
\right)
\right]\,,
\ee 
and
\be \label{eq.Omega_x}
\Omega _x  \equiv \frac{{\partial \Omega }}{{\partial x}} =  -\,\frac{B^2 x}{\Omega}
\left[
\Bigl(1 - B^2 M^2 \frac{I_2}{I_1^2}\Bigr) r^2
- 2M \frac{I_2}{I_1} r
+ a^2
\right]\,.
\ee 

The physical Maxwell potential is obtained as the real part,
\begin{equation}
A_\mu^{\rm (real)} = 2\,\mathrm{Re}\,A_\mu,
\end{equation}
so that the Einstein--Maxwell equations are satisfied with source-free electromagnetic field \(F_{\mu\nu} = \partial_\mu A^{\rm (real)}_\nu - \partial_\nu A^{\rm (real)}_\mu\). Based on the 1-form above, the non-vanishing components of the gauge potential can be written explicitly as
\be 
A_t
= \frac{1}{B\rho^2}
\left[
a\,\Omega_r\Big(a x\sqrt{1-w^2} + r w\Big)
+\frac{\Omega_x}{\sqrt{\Delta_x}}\Big(a w x - r\sqrt{1-w^2}\Big)
\right]\,,
\ee 
and 
\be 
A_\phi
= \frac{1}{B\rho^2}
\Big[
w\,(\Omega - 1)\,\rho^2
- \Omega_r\,(a^2 + r^2)\bigl(a x \sqrt{1-w^2} + r w\bigr)
+ a\,\sqrt{\Delta_x}\,\Omega_x\,\bigl(r \sqrt{1-w^2} - a w x\bigr)
\Big] \,,
\ee 
where we have used \(w=\cos\sigma\). 

In the Kerr--BR spacetime, the parameter $\sigma$ controls the duality rotation of the electromagnetic field. A purely magnetic field is achieved when $\sigma = 0$, while a purely electric field corresponds to $\sigma = \pi/2$. Importantly, the spacetime metric functions, such as $\Omega$, $P$, $\Delta$, and $Q$, are independent of $\sigma$. This means that the geometry itself is insensitive to the electric--magnetic composition of the electromagnetic field; only the electromagnetic field components vary with $\sigma$. Thus, the metric remains unchanged whether the field is purely electric, purely magnetic, or a mixture.

Several important limits follow directly from the definitions above. For vanishing external field, \(B=0\), one has
\begin{equation}
\Omega^2 = 1, \qquad
P = 1, \qquad
\rho^2 = r^2 + a^2 \cos^2\theta, \qquad
Q = \Delta = r^2 - 2 M r + a^2,
\end{equation}
and the metric reduces to the standard Kerr solution. In the opposite limit \(M \to 0\) (with \(B\neq 0\)) the geometry approaches the Bertotti--Robinson universe endowed with a uniform Maxwell field. Thus Kerr--BR provides a three-parameter family interpolating between a rotating black hole and a homogeneous \(\mathrm{AdS}_2 \times S^2\) electromagnetic background. It is also worth emphasizing that, unlike the standard Kerr black hole immersed in a Melvin universe, the external field parameter \(B\) in the Kerr--BR solution explicitly modifies both the radial function \(\Delta(r)\) and the extremality condition. As a result, the horizon locations and extremality bound are genuinely deformed by the BR embedding, whereas in the Melvin case the horizon structure of Kerr is left unchanged.

The locations of the black-hole horizons follow from the condition that the radial function \(Q(r)\) in \eqref{eq:QKBR} vanishes. Since \(Q(r) = (1 + B^2 r^2)\,\Delta(r)\), the horizons are determined by the quadratic equation \(\Delta(r)=0\), where \(\Delta\) is given in \eqref{eq:DeltaKBR}. This equation admits up to two real roots,
\begin{equation}
r_{\pm}
= I_1\,\frac{
	M I_2 \pm \sqrt{M^2 I_2 - a^2 I_1^2}
}{
	I_1^2 - B^2 M^2 I_2
} ,
\label{eq:rpm-KBR}
\end{equation}
which correspond to the outer (\(r_+\)) and inner (\(r_-\)) black-hole horizons. In the limit \(B\to 0\), where \(I_1 = I_2 = 1\), this reduces to the familiar Kerr expression
\begin{equation}
r_{\pm} = M \pm \sqrt{M^2 - a^2}.
\end{equation}
For a non-rotating black hole (\(a=0\)) one has \(I_1 = I_2 = 1\), and the quadratic equation collapses to a single positive root describing a Schwarzschild black hole in a Bertotti--Robinson universe,
\begin{equation}
r_h = \frac{2M}{1 - B^2 M^2}.
\label{eq:rh-SchBR}
\end{equation}
The horizon radius is thus shifted outward relative to the standard Schwarzschild value \(r_{\text{Schw}}=2M\); the presence of the external field effectively enlarges the black-hole horizon.

Extremal configurations occur when the two horizons merge, i.e.\ when the discriminant of \eqref{eq:rpm-KBR} vanishes:
\begin{equation}
M^2 I_2 - a^2 I_1^2 = 0.
\end{equation}
Equivalently, the extremal Kerr--BR black hole satisfies
\begin{equation}
M = \frac{a\,I_1}{\sqrt{I_2}},
\label{eq:extremal-condition}
\end{equation}
and
\be \label{eq:Mext-real}
I_2 = 1 - B^2 a^2 > 0
\ee 
so that \(I_2\) remains real and positive for a suitable range of the external field parameter \(B\) (namely \(|Ba|<1\)). For a given rotation parameter \(a\), this relation specifies the mass \(M\) at which the inner and outer horizons coalesce; conversely, for fixed \(M\) it can be regarded as determining the critical field strength at which the black hole becomes extremal. For a double root the horizon radius is simply
\begin{equation}
r_{e}
= \frac{M I_2/I_1}{1 - B^2 M^2 I_2/I_1^2}
= \frac{M}{I_1}
= \frac{a}{\sqrt{I_2}},
\label{eq:rext}
\end{equation}
where in the last equalities we have used the extremality condition \eqref{eq:extremal-condition}. In the limit \(B\to 0\), one has \(I_1 \to 1\), \(I_2 \to 1\), so that \eqref{eq:extremal-condition} reduces to \(M=a\) and \eqref{eq:rext} gives \(r_{e}=M\), recovering the usual extremal Kerr black hole. Thus the external Bertotti--Robinson field deforms both the position of the horizons and the extremality bound, yielding a two-parameter family of extremal Kerr--BR black holes labelled, for example, by \((a,B)\) or \((M,B)\).

The spacetime is axially symmetric with Killing vector \(\psi = \partial_\phi\).  The symmetry axis consists of those points where the norm of \(\psi\) vanishes, i.e.\ where \(g_{\phi\phi}=0\). In the coordinates \((t,r,x,\phi)\), this occurs at
\be 
x = \pm 1\,, \qquad (\theta = 0,\pi),
\ee 
and these two segments represent the north and south parts of the axis. To test for conical regularity, consider a small circle around the axis at fixed \(t\) and \(r\). Its circumference and radius, computed in the two-dimensional \((x,\phi)\) subspace of the metric, are
\begin{equation}
\mathcal{C}(x) = \int_0^{2\pi C} \sqrt{g_{\phi\phi}}\,d\phi,
\qquad
\mathcal{R}(x) = \int  \sqrt{g_{xx}}\,dx,
\end{equation}
where \(C\) is a constant ``conicity'' parameter specifying the range of the angular coordinate, \(\phi \in [0,2\pi C)\), and \(x_{\rm axis} = \pm 1\) denotes the value of \(x\) on the axis. For the metric \eqref{eq:KBR-metric}, one finds that near each part of the axis the ratio of circumference to radius has the finite limit
\begin{equation}
\lim_{x\to +1} \frac{\mathcal{C}}{\mathcal{R}} = 2\pi C\,P(+1),
\qquad
\lim_{x\to -1} \frac{\mathcal{C}}{\mathcal{R}} = 2\pi C\,P(-1),
\label{eq:circ-rad-limits}
\end{equation}
where \(P(x)\) is the function defined in \eqref{eq:PKBR}.  The absence of conical singularities requires that the ratio \(\mathcal{C}/\mathcal{R}\) tends to the Euclidean value \(2\pi\) on each part of the axis. Thus we demand
\begin{equation}
2\pi C\,P(+1) = 2\pi,
\qquad
2\pi C\,P(-1) = 2\pi.
\end{equation}
Since \(P(x)\) is even in \(x\), we have \(P(+1)=P(-1)\), so both conditions are satisfied by the unique choice
\begin{equation}
C = \frac{1}{P(1)}
= \Biggl[ 1 + B^2\Bigl(M^2 \frac{I_2}{I_1^2} - a^2\Bigr)\Biggr]^{-1}.
\label{eq:conicity-C}
\end{equation}
This single value of the conicity parameter regularizes the entire axis. Several limiting cases are worth noting. For vanishing external field, \(B=0\), we have \(P(1)=1\) and hence \(C=1\), recovering the usual Kerr geometry with the standard \(2\pi\)-periodic azimuthal angle. For a non-rotating Schwarzschild--Bertotti--Robinson black hole (\(a=0\)), the functions reduce to \(I_1=I_2=1\) and
\begin{equation}
C = \frac{1}{1 + B^2 M^2},
\end{equation}
which agrees with the known conicity factor required to remove conical defects on the axis of the Schwarzschild--BR spacetime.

We next summarize the basic thermodynamic properties of the Kerr--BR black hole. As usual, the entropy \(S\) and Hawking temperature \(T\) of the horizon are encoded in the horizon area \(\mathcal{A}\) and the surface gravity \(\kappa\) via
\begin{equation}
S = \frac{\mathcal{A}}{4}\,,
\qquad
T = \frac{\kappa}{2\pi}\,.
\label{eq:S-and-T}
\end{equation}
These quantities will play a central role when we test the Kerr/CFT correspondence by comparing the Bekenstein--Hawking entropy with the microscopic Cardy formula. The horizon area is obtained by integrating the angular part of the metric \eqref{eq:KBR-metric} on a spatial cross-section of the event horizon, at \(r=r_h\) and fixed \(t\). In coordinates \((x,\phi)\), this gives
\begin{equation}
\mathcal{A}(r_+)
= \int_0^{2\pi C}\!\! d\phi 
\int_{-1}^{+1} \left. \sqrt{g_{xx}\,g_{\phi\phi}} \right|_{r_ +  }  \; dx, 
\end{equation}
where \(C\) is the conicity parameter chosen such that the axis is regular. The integral can be evaluated explicitly, leading to
\begin{equation}
\mathcal{A}
= 4\pi C\,\frac{r_+^2 + a^2}{1 + B^2 r_+^2}\, .
\label{eq:Ah}
\end{equation}
For non-rotating configurations (\(a=0\)) with horizon radius \(r_h\) given in \eqref{eq:rh-SchBR}, this reduces to
\begin{equation}
\mathcal{A}
= 4\pi C\,\frac{r_{h}^2}{1 + B^2 r_{h}^2},
\end{equation}
which, after substituting the Schwarzschild--BR horizon radius \(r_h = 2M/(1 - B^2 M^2)\), reproduces the known Bertotti--Robinson result.

The surface gravity \(\kappa\) of the Killing horizon generated by the null vector \(\xi^a\) (normalized as in Kerr) may be computed in the standard way from the ``acceleration'' of \(\xi^a\),
\(\xi^b \nabla_b \xi^a = \kappa \xi^a\), or, equivalently, from the radial function \(Q(r)\). For the Kerr--BR family the usual Kerr formula applies \cite{Podolsky:2025tle}, \(\kappa = \tfrac{1}{2} Q'(r_+)/(r_+^2 + a^2)\). Using the explicit form of \(Q(r)\) and imposing the horizon condition \(Q(r_+)=0\), one can obtain
\begin{equation}
\kappa
= \frac{1 + B^2 r_+^2}{r_+^2 + a^2}
\left(
M\,\frac{I_2}{I_1}
- \frac{a^2}{r_+}
\right).
\label{eq:kappa-h}
\end{equation}
Therefore, the associated Hawking temperature follows from \eqref{eq:S-and-T},
\begin{equation}
T_H
= \frac{\kappa}{2\pi}
= \frac{1 + B^2 r_+^2}{2\pi\,(r_+^2 + a^2)}
\left(
M\,\frac{I_2}{I_1}
- \frac{a^2}{r_+}
\right).
\label{eq:TH}
\end{equation}

\begin{figure}[t]
	\centering
	\begin{subfigure}{0.48\textwidth}
		\centering
		\includegraphics[width=\textwidth]{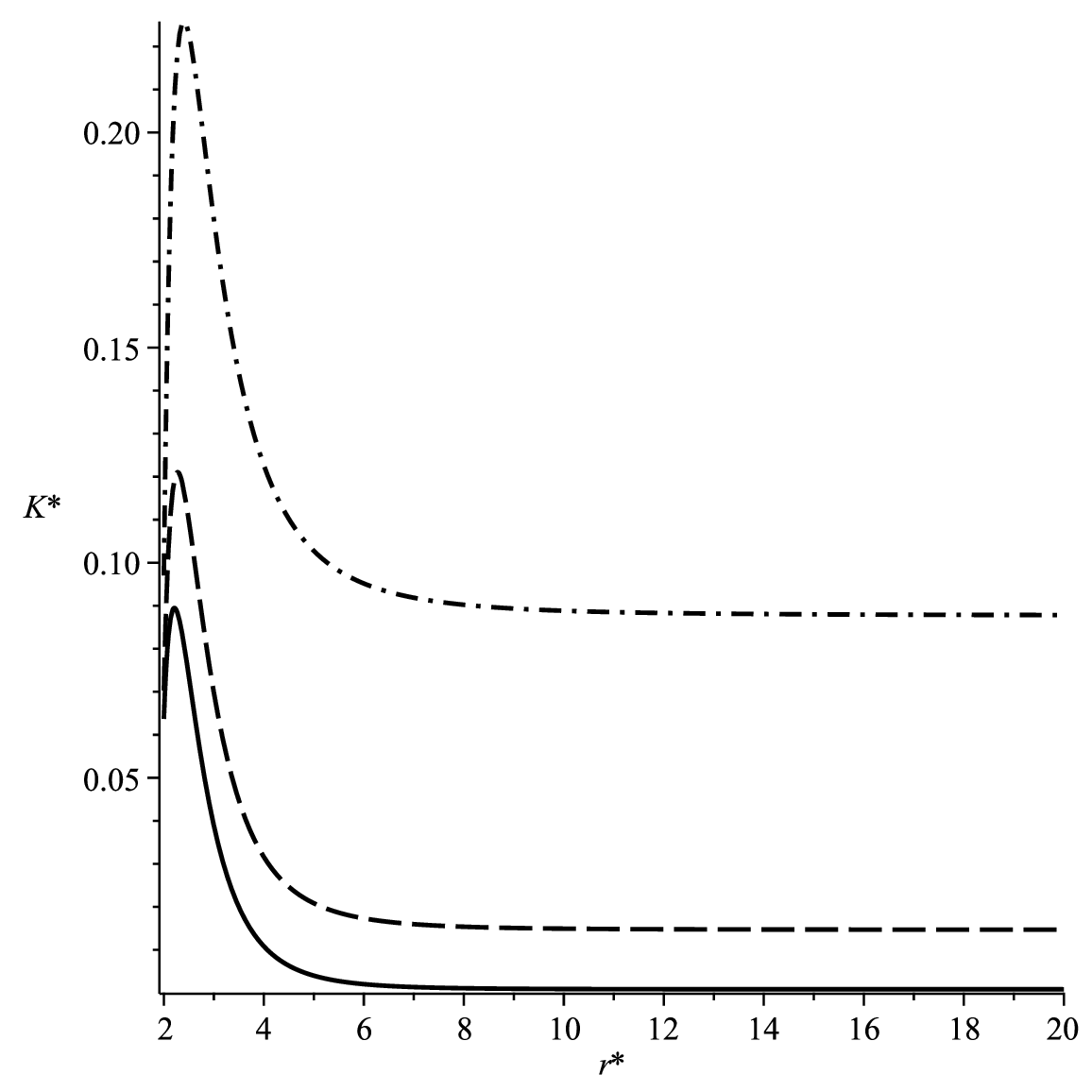}
		\caption{Dimensionless Kretschmann scalar $K^\ast$ on the equatorial plane
			as a function of the dimensionless radius $r^\ast$ in the Kerr--BR spacetime
			for three representative values of the magnetization parameter $B$.
			The solid, dashed, and dash--dotted curves correspond to increasing $|B|$.}
		\label{fig:KBR-Kfar}
	\end{subfigure}\hfill
	\begin{subfigure}{0.48\textwidth}
		\centering
		\includegraphics[width=\textwidth]{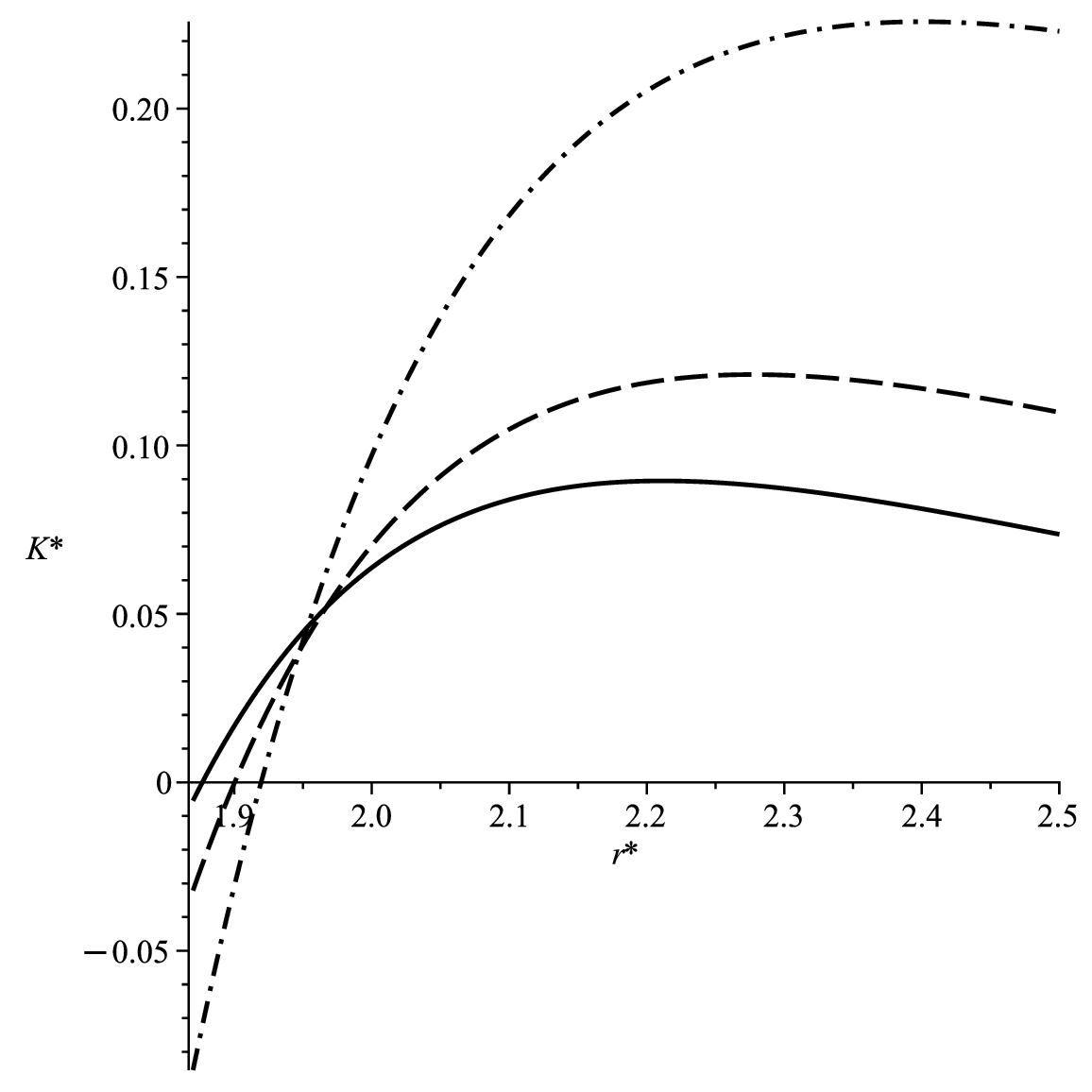}
		\caption{Zoom of panel~(a) showing the behaviour of $K^\ast$ in the near--horizon
			region $r^\ast \gtrsim r_+^\ast$ of the Kerr--BR black hole.}
		\label{fig:KBR-Knear}
	\end{subfigure}
	
	\caption{Equatorial dimensionless squared Riemann tensor
		$K^\ast = M^4 R_{\mu\nu\rho\sigma}R^{\mu\nu\rho\sigma}$ in the Kerr--BR
		spacetime for $a=0.5 M$. The dimensionless radius is $r^\ast = r/M$ and the
		outer horizon is located at $r_+^\ast \simeq 1.87$. The solid, dashed, and dash-dotted curves represent the cases $BM=0.1$, $BM=0.2$, and $BM=0.3$, respectively.}
	\label{fig:KBR-Kplots}
\end{figure}

Figure~\ref{fig:KBR-Kplots} displays the dimensionless Kretschmann scalar
$K^\ast = M^4 R_{\mu\nu\rho\sigma}R^{\mu\nu\rho\sigma}$ on the equatorial
plane of the Kerr--BR (Kerr--BR) spacetime for several values
of the magnetization parameter $B$. Panel~\ref{fig:KBR-Kfar} shows the global
radial profile: for all $B$ the curvature remains finite outside the outer
horizon at $r^\ast_+ \simeq 1.87$ and approaches a non--vanishing constant at
large $r^\ast$. This asymptotic plateau reflects the fact that the geometry is
not asymptotically flat but tends to a Bertotti--Robinson background with
constant curvature set by the external field. Increasing the magnetization
monotonically raises both the near--horizon peak and the asymptotic value of
$K^\ast$, indicating a stronger overall curvature scale.

\begin{figure}[t]
	\centering
	\begin{subfigure}{0.48\textwidth}
		\centering
		\includegraphics[width=\textwidth]{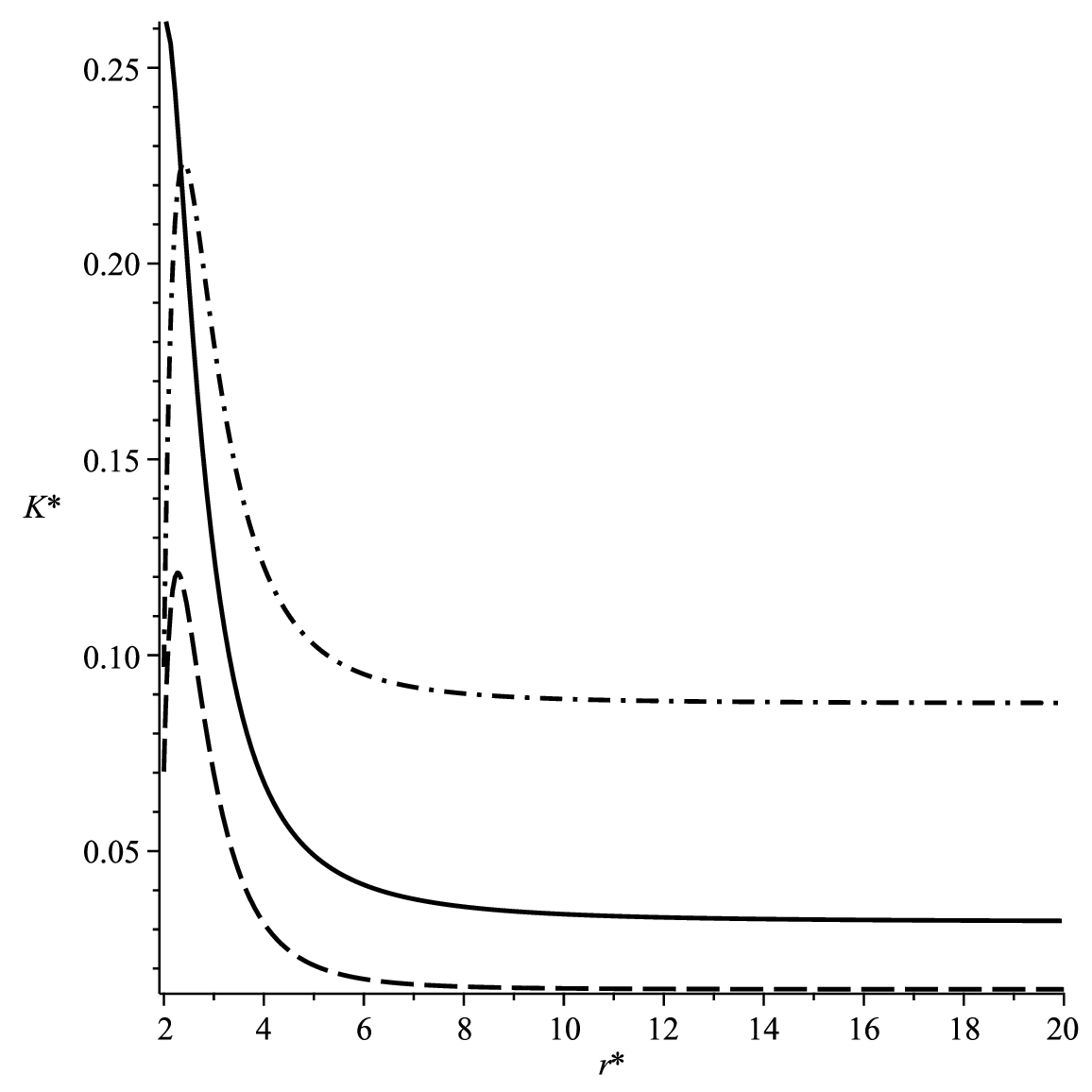}
		\caption{Equatorial dimensionless Kretschmann scalar
			$K^\ast = M^4 R_{\mu\nu\rho\sigma}R^{\mu\nu\rho\sigma}$ as a function of
			$r^\ast = r/M$ for the Melvin--Kerr spacetime. The solid, dashed and
			dash--dotted curves correspond to the same set of magnetization
			parameters $B$ as in Fig.~\ref{fig:KBR-Kplots}.}
		\label{fig:MK-Kfar}
	\end{subfigure}\hfill
	\begin{subfigure}{0.48\textwidth}
		\centering
		\includegraphics[width=\textwidth]{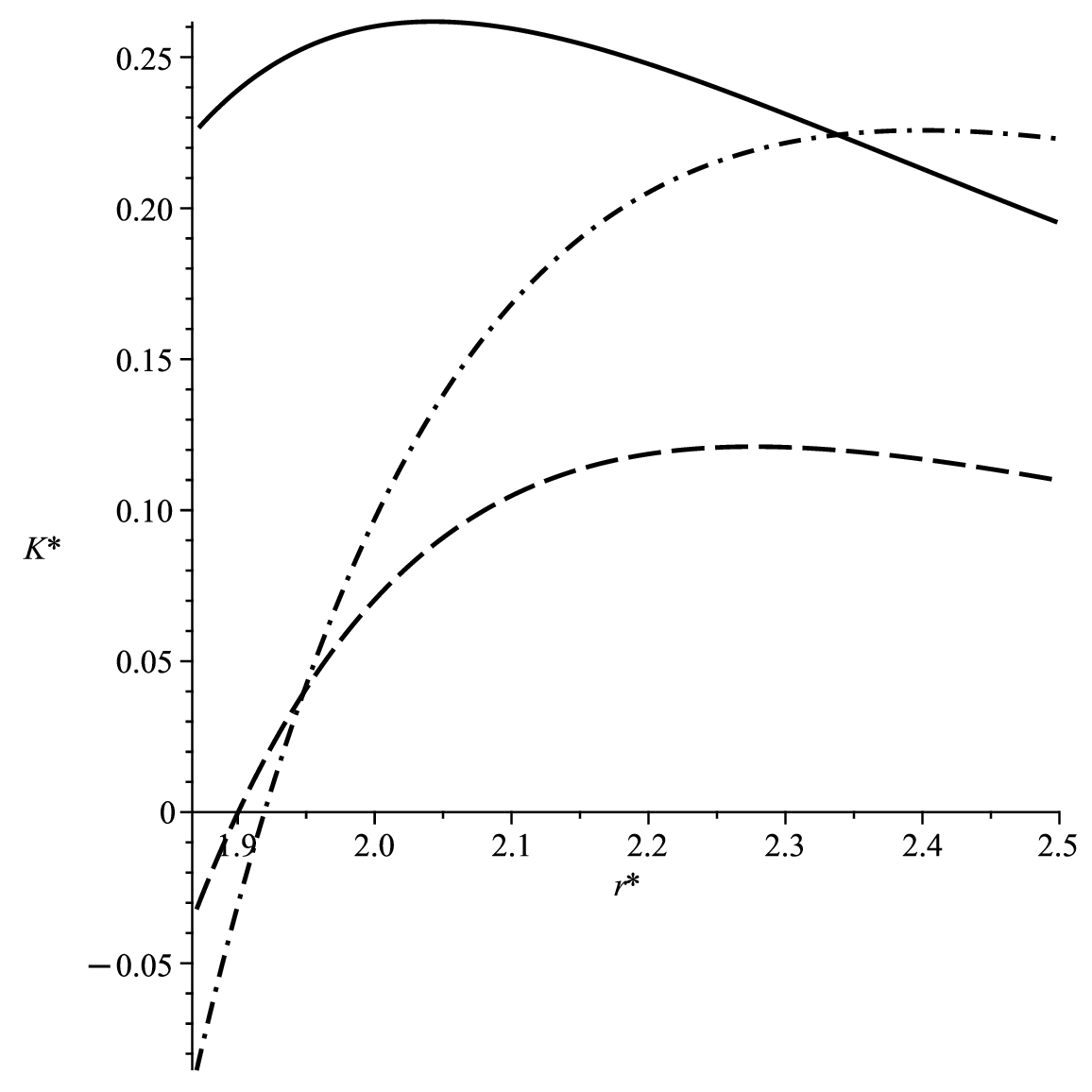}
		\caption{Near--horizon behaviour of $K^\ast$ for the Melvin--Kerr
			black hole, zoomed into the region $r^\ast \gtrsim r_+^\ast$.
			The outer horizon is again at $r_+^\ast \simeq 1.87$.}
		\label{fig:MK-Knear}
	\end{subfigure}
	
	\caption{Equatorial dimensionless squared Riemann tensor $K^\ast$ for
		the Melvin--Kerr geometry with $a=0.5$.
		The three curves represent the same magnetizations as in
		Fig.~\ref{fig:KBR-Kplots}, allowing a direct comparison between
		Kerr--BR and Melvin--Kerr backgrounds. The solid, dashed, and dash-dotted curves represent the cases $BM=0.1$, $BM=0.2$, and $BM=0.3$, respectively.}
	\label{fig:MK-Kplots}
\end{figure}

The near--horizon zoom in panel~\ref{fig:KBR-Knear} shows that, just outside
$r^\ast_+$, $K^\ast$ grows from a small value, develops a local maximum at
$r^\ast \gtrsim 2$, and then decays towards its BR plateau. For larger $B$ the
maximum becomes higher and moves slightly outward. For sufficiently strong
magnetization the curves exhibit a shallow negative dip very close to the
horizon before rising; this behaviour is not pathological, since the scalar
$R_{\mu\nu\rho\sigma}R^{\mu\nu\rho\sigma}$ in a Lorentzian spacetime is not
sign--definite. In analogy with the discussion of distorted horizons by Booth
\emph{et al.} \cite{Booth:2015nwa}, a negative value of the curvature invariant at the equator
suggests that the intrinsic geometry of the horizon two--surface may develop an
``hour--glass'' shape, with regions of negative Gaussian curvature near the
equator. As we see in Fig. \ref{fig:MK-Kplots}, an analogous feature appears in the Melvin--Kerr
case, whereas in the magnetized Kaluza--Klein background the horizon curvature
remains everywhere non--negative and such an hour--glass deformation does not
occur \cite{Siahaan:2025gid}. Overall, the plots confirm that the Kerr--BR configuration remains
regular outside $r^\ast_+$, while the external field controls both the
amplitude and the radial spread of the curvature around the black hole.

The corresponding equatorial Kretschmann profiles for the Melvin--Kerr
spacetime are shown in Fig.~\ref{fig:MK-Kplots}.  Panel~\ref{fig:MK-Kfar}
displays the global radial behaviour of $K^\ast$ for the same set of
magnetizations as in Fig.~\ref{fig:KBR-Kplots}.  As in the Kerr--BR case,
the invariant remains finite outside $r_+^\ast$ and approaches a constant
value at large $r^\ast$, now set by the Melvin background.  Increasing $B$
raises both the near--horizon peak and the asymptotic plateau, reflecting
the fact that in the Melvin--Kerr geometry the curvature at large radius is
essentially controlled by the external magnetic field rather than by the
black-hole mass.

A direct comparison of Figs.~\ref{fig:KBR-Kplots} and \ref{fig:MK-Kplots}
reveals several qualitative differences between Kerr--BR and Melvin--Kerr
curvature.  First, for a given $B$ the Melvin--Kerr profiles are more
strongly peaked near the horizon: the maximum of $K^\ast$ is higher and
located slightly closer to $r_+^\ast$ than in the Kerr--BR case, where the
profile is comparatively flatter due to the BR-type constant-curvature
background.  Second, in Kerr--BR the curvature quickly interpolates between
the horizon peak and a BR plateau whose value depends on the interplay
between rotation and magnetization, whereas in Melvin--Kerr the plateau is
predominantly set by the Melvin field and varies more monotonically with $B$.
Finally, the near-horizon zooms,
Figs.~\ref{fig:KBR-Knear} and~\ref{fig:MK-Knear}, show that for sufficiently
large $B$ the Kretschmann scalar may become slightly negative immediately
outside $r_+^\ast$ before rising to its positive maximum.  This small dip is
again consistent with the possibility of an hour--glass--like distortion of the
horizon geometry in the sense of Booth \emph{et al.} \cite{Booth:2015nwa}, now induced purely by
the external Melvin field.  In contrast, our magnetized Kaluza--Klein black
holes never develop a negative horizon curvature invariant, and their horizons
remain smoothly convex.  Overall, the comparison illustrates that the Melvin
magnetization produces stronger and more localized curvature around the
horizon, while the Kerr--BR embedding tends to spread the curvature over a
broader radial region with a lower but nonzero asymptotic value, and that both
families can exhibit nontrivial horizon shapes absent in the Kaluza--Klein
case.

At infinity, the expression of squared Riemann tensorof Kerr-BR spacetime can be shown as
\[
K_{\rm{KBR},\infty}
= \frac{8 B^{4} (B^2 a^2 - 1)^{2}}{(B^{2} a^{2} - 2)^{8}}
\Big[
(B^{2} a^{2} - 2)^{8}
- 8 B^{2} (B^{2} a^{2} - 2)^{7} M^{2}
+ 16 B^{4} (B^{2} a^{2} - 2)^{4} \big(3 B^{4} a^{4} - 8 B^{2} a^{2} + 6 \big) M^{4}
\]
\be \label{eq:KKBRinf}
- 256 B^{6} (B^2 a^2 - 1)^{2}  (B^{2} a^{2} - 2)^{2} (2 B^{2} a^{2} - 1) M^{6}
+ 1792 B^{8} (B^2 a^2 - 1)^{4} M^{8}
\Big]
\ee 
On the other hand, the similar one for Melvin-Kerr can take the form
\begin{equation}\label{eq:KMKinf}
K_{\rm{MK},\infty}
= -\,\frac{64 B^{4}\,\big(48 B^{4} M^{2} a^{2} - 5\big)}
{\big(16 B^{4} M^{2} a^{2} + 1\big)^{4}} \, .
\end{equation}
From the last equations, we learn that at spatial infinity, the squared Riemann tensor for the Kerr--BR geometry tends to the finite value \(K_{\rm KBR,\infty}\) given in Eqs.~(\ref{eq:KKBRinf}); similarly, for the Melvin--Kerr geometry it approaches \(K_{\rm MK,\infty}\) in Eq.~(\ref{eq:KMKinf}). These expressions immediately show that neither spacetime is asymptotically flat: instead of decaying as \(r\to\infty\), the curvature invariants saturate to nonzero constants controlled by the external magnetic field \(B\). In both cases the dependence on the black-hole parameters \((M,a)\) enters only through multiplicative corrections, while the overall scaling with \(B\) is fixed. A more transparent comparison is obtained by expanding both asymptotic values in powers of the magnetic field. For the Kerr--BR case one finds
\begin{equation}
K_{\rm KBR,\infty}
= 8 B^{4}
+ (32 M^{2} - 16 a^{2}) B^{6}
+ (48 M^{4} - 48 M^{2} a^{2} + 8 a^{4}) B^{8}
+ \mathcal{O}(B^{10}) \, ,
\end{equation}
whereas for the Melvin--Kerr background the expansion reads
\begin{equation}
K_{\rm MK,\infty}
= 320 B^{4}
- 23552\, M^{2} a^{2} B^{8}
+ \mathcal{O}(B^{10}) \, .
\end{equation}
In both geometries, the leading contribution is proportional to $B^{4}$ and is independent of $M$ and $a$, reflecting the fact that the far region is dominated by a Melvin-type magnetic universe, consistent with the astrophysically motivated condition $BM\ll 1$.

It is also instructive to consider special limits. In the massless limit \(M\to 0\), the Kerr--BR asymptotics reduces to
\begin{equation}
K_{\rm KBR,\infty}\big|_{M=0}
= 8 B^{4} (1 - B^{2} a^{2})^{2} \, ,
\end{equation}
which still depends on the rotation parameter through the combination \(B^{2} a^{2}\). On the other hand, for Melvin--Kerr one obtains
\begin{equation}
K_{\rm MK,\infty}\big|_{M=0}
= 320 B^{4} \, ,
\end{equation}
which coincides with the curvature of the pure Melvin universe and is completely insensitive to \(a\). This contrast emphasizes that the Kerr--BR construction retains a memory of the rotational parameter even in the absence of a mass monopole at the level of the asymptotic curvature, whereas the Melvin--Kerr family ``forgets'' about rotation in the same limit and smoothly approaches the rotationally symmetric Melvin background.

\section{Near horizon geometry of extremal Kerr--BR black holes and Cardy entropy}
\label{sec:NHEKBR}

The Kerr--BR black hole becomes extremal when the inner and outer horizons
coincide. This happens when the discriminant of \(\Delta(r)\) vanishes,
\begin{equation}
M = \frac{a\,I_1}{\sqrt{I_2}}.
\end{equation}
On this extremal branch \(\Delta(r)\) has a double root at
\begin{equation}
r_e = \frac{M}{I_1}
= \frac{a}{\sqrt{I_2}},
\label{eq:re-def}
\end{equation}
and can be written as
\begin{equation}
\Delta(r) = I_2\,(r-r_e)^2.
\label{eq:Delta-ext}
\end{equation}
Consequently,
\begin{equation}
Q(r) = (1 + B^2 r^2)\,I_2\,(r-r_e)^2,
\label{eq:Q-ext}
\end{equation}
so that the degenerate extremal horizon is located at \(r=r_e\).

On the extremal horizon the basic metric functions reduce to
\begin{equation}
\rho_0^2(x) \equiv \rho^2(r_e,x)
= r_e^2 + a^2 x^2,
\qquad
\Omega_0^2 \equiv \Omega^2(r_e,x)
= 1 + B^2 r_e^2.
\label{eq:rho0-Omega0-def}
\end{equation}
The extremality condition implies \(M^2 I_2 = a^2 I_1^2\), and therefore
\begin{equation}
P(x) = 1 + B^2\Bigl(M^2 \frac{I_2}{I_1^2} - a^2\Bigr)x^2 = 1.
\end{equation}
Thus the conicity parameter is \(C = 1/P(1)=1\) and the azimuthal coordinate has
the standard period \(2\pi\). The horizon angular velocity keeps the Kerr form
\begin{equation}
\Omega_H = \frac{a}{r_e^2 + a^2},
\label{eq:OmegaH-ext}
\end{equation}
independent of the external field \(B\).

To zoom into the near-horizon region we perform the usual Bardeen--Horowitz
scaling. Introducing
\begin{equation}
r = r_e + \lambda y,
\qquad
t = \frac{r_e^2 + a^2}{\lambda}\,\tau,
\qquad
\phi = \varphi + \Omega_H t,
\label{eq:NH-scaling}
\end{equation}
and taking \(\lambda\to 0\) with \((\tau,y,x,\varphi)\) fixed, one finds
\begin{equation}
Q(r) = (1 + B^2 r_e^2)\,I_2 (r-r_e)^2
= (1 + B^2 r_e^2)\,I_2\,\lambda^2 y^2 + \mathcal{O}(\lambda^3).
\end{equation}
Using extremality, \(r_e^2=a^2/I_2\), this coefficient becomes
\begin{equation}
(1 + B^2 r_e^2)\,I_2
= (1 + B^2 a^2/I_2)\,I_2
= I_2 + B^2 a^2 = 1,
\end{equation}
so the \((\tau,y)\)-sector already acquires the standard \(\mathrm{AdS}_2\) form
without any further rescaling.

A direct expansion of the Kerr--BR metric under the transformation
\eqref{eq:NH-scaling}, followed by linear redefinitions of \(\tau\) and
\(\varphi\) that put the \((\tau,\varphi)\) sector in canonical form, yields the
near-horizon metric (NHEK--BR)
\begin{equation}
\begin{aligned}
ds^2_{\rm nh}
={}&
\frac{\rho_0^2(x)}{\Omega_0^2}
\left(
- y^2 d\tau^2 + \frac{dy^2}{y^2}
+ \frac{dx^2}{1-x^2}
\right)
\\[2pt]
&\quad
+ \frac{(1-x^2)}{\Omega_0^2\,\rho_0^2(x)}
\big(r_e^2 + a^2\big)^2
\left(d\varphi + k\,y\,d\tau\right)^2 .
\end{aligned}
\label{eq:NHEKBR-metric}
\end{equation}
The twist parameter controlling the fibration of the azimuthal circle over the
\(\mathrm{AdS}_2\) base is
\begin{equation}
k
= \,\frac{2\sqrt{1-B^2 a^2}}{2-B^2 a^2},
\label{eq:kappa-def}
\end{equation}
which smoothly tends to \(k\to 1\) as \(B\to 0\), recovering the
standard NHEK geometry. In the same limit \(\Omega_0^2\to 1\) and \(r_e\to M=a\),
so \eqref{eq:NHEKBR-metric} reduces to the familiar NHEK metric
\cite{Guica:2008mu,Compere:2012jk}. 

It is convenient to rewrite \eqref{eq:NHEKBR-metric} in the canonical Kerr/CFT
form
\begin{equation}
ds^2_{\rm nh}
= \Gamma(x)\left(
- y^2 d\tau^2 + \frac{dy^2}{y^2} + \alpha(x)^2\,dx^2
\right)
+ \gamma(x)^2\,(d\varphi + k y\,d\tau)^2,
\label{eq:NHEKBR-canonical}
\end{equation}
with
\begin{equation}
\Gamma(x) = \frac{\rho_0^2(x)}{\Omega_0^2},
\qquad
\alpha(x)^2 = \frac{1}{1-x^2},
\qquad
\gamma(x)^2 = \frac{(1-x^2)}{\Omega_0^2\,\rho_0^2(x)}
\big(r_e^2 + a^2\big)^2.
\label{eq:Gamma-alpha-gamma}
\end{equation}
The isometry group of \eqref{eq:NHEKBR-canonical} is
\(SL(2,\mathbb{R})\times U(1)\), generated by the standard \(\mathrm{AdS}_2\)
Killing vectors in the \((\tau,y)\) sector and the rotational Killing vector
\(\partial_\varphi\).

The electromagnetic sector admits an equally simple near-horizon description.
In a gauge regular on the future horizon, the near-horizon gauge field can be
written in the compact form
\begin{equation}
{\bf A}_{\rm nh}
= {\cal Z}\,\big(d\varphi + k y\,d\tau\big)\,,
\label{eq:NHEKBR-gauge}
\end{equation}
with
\begin{equation}
{\cal Z}
= \,\frac{a^2 B\,(2 - B^2 a^2)}{2\sqrt{1-B^2 a^2}}\,.
\label{eq:Zs-def}
\end{equation}
The gauge field is manifestly aligned with the fibration appearing in the metric
\eqref{eq:NHEKBR-canonical}. The constant term proportional to \(d\varphi\) is
locally pure gauge; the physical content resides in the field strength
\({\bf F}_{\rm nh}={\rm d}{\bf A}_{\rm nh}\). Note that we restrict to the parameter range $B^2 a^2<1$ so that $k$ and ${\cal Z}$ are real and finite. The extremal horizon area follows from the general expression evaluated at
\(r=r_e\) with \(C=1\),
\begin{equation}
\mathcal{A}_{\rm ext}
= 4\pi\,\frac{r_e^2 + a^2}{1 + B^2 r_e^2},
\label{eq:Aext}
\end{equation}
so that the Bekenstein--Hawking entropy is
\begin{equation}
S_{\rm BH}
= \frac{\mathcal{A}_{\rm ext}}{4}
= \pi\,\frac{r_e^2 + a^2}{1 + B^2 r_e^2}.
\label{eq:SBH}
\end{equation}

The near-horizon metric (\eqref{eq:NHEKBR-canonical}) and vector potential (\ref{eq:NHEKBR-gauge}) have exactly the structure
used in Kerr/CFT analyses \cite{Guica:2008mu,Compere:2012jk} and in the
accelerating Kerr--Taub--NUT case reviewed in \cite{Siahaan:2024zhx}. Following
the prescription of these works, we impose the usual Kerr/CFT boundary
conditions on perturbations of the metric and gauge field. The diffeomorphisms
generated by \cite{Hartman:2008pb,Compere:2012jk}
\begin{equation}
\zeta_n = -e^{-in\varphi}\partial_\varphi
- in e^{-in\varphi} y\partial_y
+ \cdots
\end{equation}
span a single copy of the Virasoro algebra associated with the \(U(1)\)
rotational symmetry. Using the covariant phase space formalism \cite{Compere:2012jk}, the corresponding
conserved charges obey a Virasoro algebra with central charge
\begin{equation}
c_L
= 3 \kappa\!\int_{-1}^{+1}\!\!dx\;
\sqrt{\Gamma(x)\,\gamma(x)^2\,\alpha(x)^2},
\label{eq:cL-int}
\end{equation}
which is the analogue of the expression used in
\cite{Hartman:2008pb,Compere:2012jk}. For the metric
\eqref{eq:NHEKBR-canonical} this integral is simply proportional to the extremal
horizon area, yielding
\begin{equation}
c_L = \frac{3\kappa}{2\pi}\,\mathcal{A}_{\rm ext}
= 12\,\frac{\sqrt{1-B^2 a^2}}{2 - B^2 a^2}\,
\frac{r_e^2 + a^2}{1 + B^2 r_e^2}.
\label{eq:cL-raw}
\end{equation}
Introducing an effective angular momentum
\begin{equation}
{\cal J} \equiv \frac{1}{2}\,\frac{r_e^2 + a^2}{1 + B^2 r_e^2},
\end{equation}
this can be written succinctly as
\begin{equation}
c_L = 12\,\kappa\,{\cal J},
\end{equation}
so the external Bertotti--Robinson field enters through the rescaled effective
angular momentum \({\cal J}\) and the twist \(\kappa\). In the Kerr limit
\(B\to 0\) one has \({\cal J}\to J=aM\) and \(\kappa\to 1\), recovering
\(c_L=12J\) in the original Kerr/CFT correspondence proposal \cite{Guica:2008mu}.

The Frolov--Thorne temperature of the left-moving sector of the dual CFT is
obtained, as in \cite{Guica:2008mu,Hartman:2008pb,Compere:2012jk}, from the
extremal limit of the Hartle--Hawking vacuum,
\begin{equation}
T_L = \frac{1}{2\pi \kappa},
\label{eq:TL}
\end{equation}
so the dependence on the external field is entirely encoded in \(\kappa(a,B)\).
The microscopic entropy of the dual chiral CFT follows from the Cardy formula,
\begin{equation}
S_{\rm CFT}
= \frac{\pi^2}{3}\,c_L\,T_L.
\end{equation}
Substituting \eqref{eq:cL-raw} and \eqref{eq:TL} gives
\begin{equation}
S_{\rm CFT}
= \frac{\pi^2}{3}
\left(6 \kappa\,\frac{r_e^2 + a^2}{1 + B^2 r_e^2}\right)
\frac{1}{2\pi \kappa}
= \pi\,\frac{r_e^2 + a^2}{1 + B^2 r_e^2}.
\end{equation}
Comparing with \eqref{eq:SBH}, we find
\begin{equation}
S_{\rm CFT} = S_{\rm BH} = \frac{\mathcal{A}_{\rm ext}}{4}.
\end{equation}
Thus the Kerr/CFT correspondence holds for extremal Kerr--BR
black holes, namely the microscopic entropy of the dual chiral CFT exactly reproduces
the Bekenstein--Hawking entropy of the extremal black holes embedded in any type of
external Bertotti--Robinson field.

\section{Conclusion}\label{sec:Conclusion}

In this work, we have analyzed the geometry and holographic properties of the Kerr--Bertotti--Robinson (Kerr--BR) black hole, an exact Einstein--Maxwell solution describing a rotating black hole immersed in a homogeneous Bertotti--Robinson electromagnetic universe. By reviewing the metric, electromagnetic potential, horizon structure, and thermodynamics, we demonstrated how the external BR field explicitly shifts both the horizon locations and the extremality condition. It is one of the distinct features compared to the Melvin--Kerr spacetime, where the magnetization leaves the Kerr bound unchanged. A robust diagnostic of these differences is provided by the equatorial Kretschmann scalar: while the dimensionless profiles remain finite outside the horizon and approach a nonzero constant at large radius for both geometries, their detailed radial behavior and the onset of "hour-glass" horizon deformations differ qualitatively between the Kerr--BR, Melvin--Kerr, and magnetized Kaluza--Klein cases.

The principal result of this paper is the establishment of a Kerr/CFT correspondence for extremal Kerr--BR black holes. On the extremal branch, the Bardeen--Horowitz scaling reveals a near-horizon metric of warped \(\mathrm{AdS}_3\) form, characterized by functions \(\Gamma(x)\), \(\alpha(x)\), \(\gamma(x)\) and a twist parameter \(\kappa(a,B)\) that encodes the external field's influence. The associated near-horizon Maxwell field is naturally aligned with the \(U(1)\) fiber and takes the form \({\bf A}_{\rm nh} = {\cal Z}\,(d\varphi + \kappa y\,d\tau)\). Upon imposing standard Kerr/CFT boundary conditions, the asymptotic symmetry algebra yields a single Virasoro copy with central charge \(c_L = 12\,\kappa\,{\cal J}\), where \({\cal J}\) is an effective angular momentum proportional to the extremal horizon area. Combining this with the left-moving Frolov--Thorne temperature, \(T_L = 1/(2\pi\kappa)\), the microscopic entropy computed via the Cardy formula, \(S_{\rm CFT} = S_{\rm BH} = \mathcal{A}_{\rm ext}/4\), exactly reproduces the macroscopic Bekenstein--Hawking entropy for arbitrary values of the external field compatible with extremality. This confirms that the Kerr/CFT paradigm robustly survives the immersion of a rotating black hole into a homogeneous \(\mathrm{AdS}_2\times S^2\) electromagnetic universe.

Several possibilities for future research naturally arise. On the gravity side, it would be valuable to investigate quasi-normal modes, stability, and low-frequency scattering of test fields in the Kerr--BR background. A systematic comparison with Melvin--Kerr and magnetized Kaluza--Klein black holes, specifically regarding horizon geometry and Gaussian curvature, could offer deeper insights into strong-field magnetospheric effects. On the holographic side, extending the Kerr/CFT correspondence to the more general class of black holes embedded in Bertotti--Robinson--Bonnor--Melvin fields \cite{Astorino:2025lih} constitutes a promising direction for future investigation.

\section*{Acknowledgement}

This work was supported by LPPM-UNPAR through the Penelitian Publikasi Internasional Bereputasi funding scheme.

\appendix

\section{Equatorial squared Riemann tensor}
\label{app:equatorial-K}

On the equatorial plane \((\theta = \pi/2)\), the squared Riemann tensor
(Kretschmann scalar) for the Kerr--BR spacetime can be written
in the compact form
\begin{equation} \label{eq:RRkbr}
K_{\text{KBR}}(r)
= \frac{8\,(B^2 a^2 - 1)^2}{\bigl(B^2 a^2 - 2\bigr)^8 \bigl(a^2 + r^2\bigr)^6}
\sum_{j = 0}^{12} k_j\, r^j ,
\end{equation}
where the coefficients \(k_j\) are given by
\begin{align}
k_{12}
&= B^{4}(B^{2}a^{2}-2)^{8}
- 8\,B^{6}(B^{2}a^{2}-2)^{7} M^{2} 
+ 16\,B^{8}(B^{2}a^{2}-2)^{4}
\bigl(3 B^{4}a^{4} - 8 B^{2}a^{2} + 6\bigr) M^{4} \notag\\
&\quad
- 256\,B^{10}(B^{2}a^{2}-1)^{2}(B^{2}a^{2}-2)^{2}(2 B^{2}a^{2}-1)\, M^{6} 
+ 1792\,B^{12}(B^{2}a^{2}-1)^{4} M^{8} \, .
\end{align}
\begin{align}
k_{11}
&= -64\,B^{8} M^{3}\,(B^{2} a^{2} - 1)(B^{2} a^{2} - 2)\Bigl[
8 B^{2} (B^{2} a^{2} - 1)^{2} (23 B^{2} a^{2} - 9)\,M^{4} \notag\\
&\qquad\qquad\qquad
- 4 B^{2} a^{2} (B^{2} a^{2} - 2)^{2} (4 B^{2} a^{2} - 5)\,M^{2}
+ a^{2} (B^{2} a^{2} - 2)^{4}
\Bigr].
\end{align}
\begin{align}
k_{10}
&= -2 B^{4}\Bigl[
4352\,B^{8} a^{2}\,(B^{2} a^{2} - 1)^{4} M^{8}
- 192\,B^{4}\,(B^{2} a^{2} - 2)^{2} (B^{2} a^{2} - 1)^{2}
\bigl(65 B^{4} a^{4} - 72 B^{2} a^{2} + 15\bigr) M^{6} \notag\\
&\qquad
+ 16\,B^{4} a^{2}\,(B^{2} a^{2} - 2)^{4}
\bigl(19 B^{4} a^{4} - 40 B^{2} a^{2} + 18\bigr) M^{4} \notag\\
&\qquad
- 8\,B^{2} a^{2}\,(B^{2} a^{2} - 2)^{6} (B^{2} a^{2} + 2)\,M^{2}
- 3\,a^{2}\,(B^{2} a^{2} - 2)^{8}
\Bigr].
\end{align}
\begin{align}
k_9
&= 16\,B^{4} M\,(B^{2} a^{2} - 1)(B^{2} a^{2} - 2)\Bigl[
1984\,B^{12} M^{6} a^{8}
- 1568\,B^{12} M^{4} a^{10}
- B^{12} a^{14} \notag\\[2pt]
&\qquad
- 6080\,B^{10} M^{6} a^{6}
+ 9328\,B^{10} M^{4} a^{8}
- 20\,B^{10} M^{2} a^{10}
+ 12\,B^{10} a^{12} \notag\\[2pt]
&\qquad
+ 6208\,B^{8} M^{6} a^{4}
- 20288\,B^{8} M^{4} a^{6}
+ 160\,B^{8} M^{2} a^{8}
- 60\,B^{8} a^{10} \notag\\[2pt]
&\qquad
- 2112\,B^{6} M^{6} a^{2}
+ 19632\,B^{6} M^{4} a^{4}
- 480\,B^{6} M^{2} a^{6}
+ 160\,B^{6} a^{8} \notag\\[2pt]
&\qquad
- 8128\,B^{4} M^{4} a^{2}
+ 640\,B^{4} M^{2} a^{4}
- 240\,B^{4} a^{6} \notag\\[2pt]
&\qquad
+ 960\,B^{2} M^{4}
- 320\,B^{2} M^{2} a^{2}
+ 192\,B^{2} a^{4}
- 64\,a^{2}
\Bigr].
\end{align}
\begin{align}
k_8
&= B^{4}\Bigl[
B^{16}\bigl(
1792\,M^{8} a^{12}
- 46848\,M^{6} a^{14}
+ 13488\,M^{4} a^{16}
+ 120\,M^{2} a^{18}
+ 15\,a^{20}
\bigr) \notag\\[2pt]
&\quad
+ B^{14}\bigl(
384768\,M^{6} a^{12}
- 7168\,M^{8} a^{10}
- 145984\,M^{4} a^{14}
- 1424\,M^{2} a^{16}
- 240\,a^{18}
\bigr) \notag\\[2pt]
&\quad
+ B^{12}\bigl(
1281280\,M^{6} a^{10}
+ 10752\,M^{8} a^{8}
+ 666272\,M^{4} a^{12}
+ 6992\,M^{2} a^{14}
+ 1680\,a^{16}
\bigr) \notag\\[2pt]
&\quad
+ B^{10}\bigl(
2211072\,M^{6} a^{8}
- 7168\,M^{8} a^{6}
- 1663424\,M^{4} a^{10}
- 18048\,M^{2} a^{12}
- 6720\,a^{14}
\bigr) \notag\\[2pt]
&\quad
+ B^{8}\bigl(
1792\,M^{8} a^{4}
- 2083840\,M^{6} a^{6}
+ 2462112\,M^{4} a^{8}
+ 25280\,M^{2} a^{10}
+ 16800\,a^{12}
\bigr) \notag\\[2pt]
&\quad
+ B^{6}\bigl(
1016832\,M^{6} a^{4}
- 2182912\,M^{4} a^{6}
- 16640\,M^{2} a^{8}
- 26880\,a^{10}
\bigr) \notag\\[2pt]
&\quad
+ B^{4}\bigl(
1106176\,M^{4} a^{4}
- 200704\,M^{6} a^{2}
+ 768\,M^{2} a^{6}
+ 26880\,a^{8}
\bigr) \notag\\[2pt]
&\quad
+ B^{2}\bigl(
4096\,M^{2} a^{4}
- 278528\,M^{4} a^{2}
- 15360\,a^{6}
\bigr) \notag\\[2pt]
&\quad
+ 23040\,M^{4}
- 1024\,M^{2} a^{2}
+ 3840\,a^{4}
\Bigr].
\end{align}
\begin{align}
k_7
&= -32\,B^{2} M\,(B^{2} a^{2} - 1)(B^{2} a^{2} - 2)\Bigl[
16 B^{8} a^{4} (B^{2} a^{2} - 1)^{2}(11 B^{2} a^{2} - 21)\,M^{6} \notag\\
&\qquad\qquad\qquad
- 8 B^{2} (B^{2} a^{2}) (B^{2} a^{2} - 2)^{2}
\bigl(20 B^{2} a^{2} - 17\bigr)
\bigl(7 (B^{2} a^{2})^{2} - 18 B^{2} a^{2} + 9\bigr)\,M^{4} \notag\\
&\qquad\qquad\qquad
+ (B^{2} a^{2} - 2)^{4}
\bigl(125 (B^{2} a^{2})^{3}
- 195 (B^{2} a^{2})^{2}
+ 99 B^{2} a^{2}
- 9\bigr)\,M^{2} \notag\\
&\qquad\qquad\qquad
+ 2 B^{2} a^{4}(B^{2} a^{2} - 2)^{6}
\Bigr].
\end{align}
\begin{align}
k_6
&= 4\,(B^{2} a^{2} - 2)^{2}\Bigl[
32\,B^{8} a^{4}\,(B^{2} a^{2} - 1)^{2}
\bigl(59 (B^{2} a^{2})^{2} - 220 B^{2} a^{2} + 185\bigr)\,M^{6} \notag\\
&\qquad
- 16\,B^{4} a^{2}\,(B^{2} a^{2} - 2)^{2}
\bigl(231 (B^{2} a^{2})^{4} - 1094 (B^{2} a^{2})^{3}
+ 1740 (B^{2} a^{2})^{2} - 1146 B^{2} a^{2} + 264\bigr)\,M^{4} \notag\\
&\qquad
+ 2\,(B^{2} a^{2} - 2)^{4}
\bigl(85 (B^{2} a^{2})^{4} - 158 (B^{2} a^{2})^{3}
+ 132 (B^{2} a^{2})^{2} - 42 B^{2} a^{2} + 3\bigr)\,M^{2} \notag\\
&\qquad
+ 5\,B^{4} a^{6}\,(B^{2} a^{2} - 2)^{6}
\Bigr].
\end{align}
\begin{align}
k_5
&= -32\,M B^{2} a^{2}\,(B^{2} a^{2} - 1)(B^{2} a^{2} - 2)^{3}\Bigl[
8 B^{2} (B^{2} a^{2}) \bigl(22 (B^{2} a^{2})^{3} - 117 (B^{2} a^{2})^{2}
+ 192 B^{2} a^{2} - 93\bigr) M^{4} \notag\\
&\qquad
- 2 (B^{2} a^{2} - 2)^{2}
\bigl(45 (B^{2} a^{2})^{3} - 204 (B^{2} a^{2})^{2}
+ 209 B^{2} a^{2} - 60\bigr) M^{2} \notag\\
&\qquad
+ 3 a^{2} (B^{2} a^{2}) (B^{2} a^{2} - 2)^{4}
\Bigr].
\end{align}
\begin{align}
k_4
&= a^{2}(B^{2} a^{2} - 2)^{2}\Bigl[
-256\,B^{8}a^{4}\,(B^{2}a^{2} - 1)^{2} M^{6} \notag\\
&\qquad
+ 16\,B^{2}(B^{2}a^{2})(B^{2}a^{2} - 2)^{2}
\bigl(160 (B^{2}a^{2})^{4} - 1028 (B^{2}a^{2})^{3}
+ 2412 (B^{2}a^{2})^{2} - 2316 B^{2}a^{2} + 787\bigr) M^{4} \notag\\
&\qquad
- 8\,(B^{2}a^{2} - 2)^{4}
\bigl(6 (B^{2}a^{2})^{4} - 164 (B^{2}a^{2})^{3}
+ 308 (B^{2}a^{2})^{2} - 210 B^{2}a^{2} + 45\bigr) M^{2} \notag\\
&\qquad
+ 15\,a^{2}(B^{2}a^{2})^{2}(B^{2}a^{2} - 2)^{6}
\Bigr].
\end{align}
\begin{align}
k_3
&= -32\,B^{2} M a^{4}\,(B^{2} a^{2} - 1)(B^{2} a^{2} - 2)^{3}\Bigl[
8 B^{4} a^{2}\bigl(3 - 2 B^{2} a^{2}\bigr)\,M^{4} \notag\\
&\qquad\qquad
+ (B^{2} a^{2} - 2)^{2}
\bigl(23 (B^{2} a^{2})^{3}
- 107 (B^{2} a^{2})^{2}
+ 199 B^{2} a^{2} - 105\bigr)\,M^{2} \notag\\
&\qquad\qquad
+ 2 B^{2} a^{4}(B^{2} a^{2} - 2)^{4}
\Bigr].
\end{align}
\begin{align}
k_2
&= 2\,a^{4}(B^{2} a^{2} - 2)^{4}\Bigl[
16 B^{2} (B^{2} a^{2})\bigl(-12 (B^{2} a^{2})^{2}
+ 32 B^{2} a^{2} - 17\bigr) M^{4} \notag\\
&\qquad\qquad
+ 4 (B^{2} a^{2} - 2)^{2}
\bigl(17 (B^{2} a^{2})^{4} - 62 (B^{2} a^{2})^{3}
+ 132 (B^{2} a^{2})^{2} - 126 B^{2} a^{2} + 45\bigr) M^{2} \notag\\
&\qquad\qquad
+ 3 a^{2} (B^{2} a^{2})^{2} (B^{2} a^{2} - 2)^{4}
\Bigr].
\end{align}
\begin{align}
k_1
&= -16\,B^{2} M a^{6}\,(B^{2} a^{2} - 1)(B^{2} a^{2} - 2)^{5}\Bigl[
B^{6} a^{8}
- 4\,B^{4} a^{6}
- 8\,B^{2} M^{2} a^{2}
+ 4\,B^{2} a^{4}
+ 12\,M^{2}
\Bigr],
\end{align}
and
\begin{align}
k_0
&= a^{6}(B^{2} a^{2} - 2)^{4}\Bigl[
16 B^{4} a^{2} M^{4}
- 8\,(B^{2} a^{2} - 2)^{2}\bigl(2 B^{4} a^{4} - 6 B^{2} a^{2} + 3\bigr) M^{2}
+ B^{4} a^{6}(B^{2} a^{2} - 2)^{4}
\Bigr].
\end{align}

For comparison, the equatorial squared Riemann tensor in the Melvin--Kerr
spacetime takes the form
\begin{align}\label{eq:RRmk}
K_{\text{MK}}(r)
&= -\frac{16}{(a^{2}+r^{2})^{6}\,\bigl(16 B^{4} M^{2} a^{2} + 1\bigr)^{4}}
\Bigl[
B^{8}\bigl(
768 M^{6} a^{10}
- 11520 M^{6} a^{8} r^{2}
+ 11520 M^{6} a^{6} r^{4}
\notag\\[2pt]
&\qquad
- 768 M^{6} a^{4} r^{6}
+ 768 M^{4} a^{12}
- 4608 M^{4} a^{8} r^{4}
- 6144 M^{4} a^{6} r^{6}
- 2304 M^{4} a^{4} r^{8} \notag\\[2pt]
&\qquad
+ 192 M^{2} a^{14}
+ 1152 M^{2} a^{12} r^{2}
+ 2880 M^{2} a^{10} r^{4}
+ 3840 M^{2} a^{8} r^{6} \notag\\[2pt]
&\qquad
+ 2880 M^{2} a^{6} r^{8}
+ 1152 M^{2} a^{4} r^{10}
+ 192 M^{2} a^{2} r^{12}
\bigr) \notag\\[4pt]
&\quad
+ B^{6}\bigl(
576 M^{3} a^{10} r
+ 1536 M^{3} a^{8} r^{3}
+ 1152 M^{3} a^{6} r^{5}
- 192 M^{3} a^{2} r^{9}
\bigr) \notag\\[4pt]
&\quad
+ B^{4}\bigl(
96 M^{4} a^{8}
- 1440 M^{4} a^{6} r^{2}
+ 1440 M^{4} a^{4} r^{4}
- 96 M^{4} a^{2} r^{6}
+ 48 M^{2} a^{10}
- 288 M^{2} a^{6} r^{4} \notag\\[2pt]
&\qquad
- 384 M^{2} a^{4} r^{6}
- 144 M^{2} a^{2} r^{8}
- 20 a^{12}
- 120 a^{10} r^{2}
- 300 a^{8} r^{4}
- 400 a^{6} r^{6}  \notag\\[2pt]
&\qquad
- 300 a^{4} r^{8}
- 120 a^{2} r^{10}
- 20 r^{12}
\bigr) \notag\\[4pt]
&\quad
+ B^{2}\bigl(
36 M a^{8} r
+ 96 M a^{6} r^{3}
+ 72 M a^{4} r^{5}
- 12 M r^{9}
\bigr) \notag\\[4pt]
&\quad
+ 3 M^{2} a^{6}
- 45 M^{2} a^{4} r^{2}
+ 45 M^{2} a^{2} r^{4}
- 3 M^{2} r^{6}
\Bigr].	
\end{align}

\end{document}